\documentclass[10pt,oneside]{article} 
\usepackage[margin=1in]{geometry}    
\usepackage[utf8]{inputenc}
\usepackage[T1]{fontenc}
\usepackage{authblk}                 
\usepackage{setspace}                
\usepackage{lineno}
\usepackage{hyperref}                
\usepackage{graphicx}
\usepackage{cite}

\title{\textbf{Fiber Link Stabilization with a Multicore Fiber Amplifier}}

\author[1,2,*]{Yifan Liu}
\author[2,3]{Takuma Nakamura}
\author[4]{Daniel J. Elson}
\author[4]{Yuta Wakayama}
\author[2]{Charles A. McLemore}
\author[5]{Tetsuya Hayashi}
\author[6]{Mikael Mazur}
\author[6]{Nicolas Fontaine}
\author[2,7]{Franklyn Quinlan}
\author[2,$\dagger$]{Nazanin Hoghooghi}

\affil[1]{Department of Physics, University of Colorado Boulder, 440 UCB Boulder, CO 80309, USA}
\affil[2]{Time and Frequency Division, National Institute of Standards and Technology, Boulder, CO 80305, USA}
\affil[3]{Department of Electrical Engineering, University of Colorado Denver, 1200 Larimer Street, Denver, CO 80204, USA}
\affil[4]{KDDI Research, Inc., 2-1-15 Ohara, Fujimino 356-8502, Japan}
\affil[5]{Sumitomo Electric Industries, Ltd., 1, Taya-cho, Sakae-ku, Yokohama, Kanagawa, Japan}
\affil[6]{Nokia Bell Labs, Murray Hill, New Providence, NJ 07974, USA}
\affil[7]{Electrical, Computer and Energy Engineering, University of Colorado, Boulder, Colorado 80309, USA}

\affil[*]{yifan.liu@colorado.edu}
\affil[$\dagger$]{nazanin.hoghooghi@nist.gov}
\date{} 

\begin{document}
\maketitle

\begin{abstract}
We study the use of separate cores of a multicore erbium-doped fiber amplifier (MC-EDFA) in a noise-canceled link for ultrastable optical frequency transfer. We demonstrate fractional frequency instability of $5\times10^{-19}$ at 1000~s averaging time for the stabilized MC-EDFA alone and $1.4\times10^{-18}$ at 1000~s averaging time when integrated with a 40 km-long 7-core spooled fiber. This study further establishes multicore fiber (MCF) networks as a promising platform for ultrastable frequency transfer, serving as an important step toward incorporating precision time and frequency distribution into future MCF communication infrastructures.
\end{abstract}

\maketitle

\section{Introduction}
The vast global network of single-core (SC) optical fibers deployed by the telecommunications industry can be leveraged for additional applications, such as environmental sensing and ultrastable optical frequency transfer \cite{Marra2018-dp,Lisdat2016-jf,Predehl2012-df}. As these fiber networks are designed and built for data transfer, any ancillary use must be compatible with the routing infrastructure for low error-rate signals. In the case of ultrastable optical frequency transfer, the main challenge lies in creating long-haul phase-stabilized links. Phase stabilization requires bidirectional operation of the link to measure and compensate for environmental perturbations that disturb the phase of the light and degrade signal stability \cite{Ma1994-lt}. However, SC fiber networks typically operate unidirectionally. As an important example, optical amplifiers in telecom links contain isolators to mitigate back-reflections and spurious lasing. Ultrastable frequency transfer over such links, therefore, requires the installation of custom bidirectional amplifiers. While links with SC bidirectional amplifiers have enabled intra-continental state-of-the-art optical clock comparisons \cite{Schioppo2022-xv}, difficulties with back-reflections remain, such that the gain of each amplifier must be severely curtailed \cite{Predehl2012-df}. Moreover, reconfiguring a link with bidirectional amplifiers is only practical for land-based links where access to the amplifiers is feasible.

Currently, fiber networks based on weakly-coupled multicore fibers (MCFs) are being actively investigated. These MCFs create an exciting possibility for ultrastable optical signal distribution to coexist with telecom data traffic. Their key advantage is that the multiple cores share the same cladding, leading to highly correlated phase fluctuations between them—much more so than between SC fibers bundled together. Therefore, a noise-canceled link can be implemented by sending the ultrastable signal through one core and returning a portion of it through another, generating the signal needed for noise compensation while maintaining unidirectional propagation in each core. Using this method, a 25~km deployed 4-core fiber link was stabilized to a level that can support the transfer of state-of-the-art optical clock signals \cite{Hoghooghi2025-st}. However, establishing the full capability of MCF links for ultrastable optical signal transfer requires integrating erbium-doped fiber amplifiers (EDFAs) into the link. So far, no study has demonstrated whether multiple cores of a multicore EDFA (MC-EDFA) can maintain the necessary phase correlation to support high-fidelity frequency transfer.

Here, we investigate the compatibility of a MC-EDFA for ultrastable frequency transfer. By integrating an MC-EDFA with a 40-km, 7-core MCF, we demonstrate a stabilized link that achieves a fractional frequency instability of $1.4\times 10^{-18}$ at 1000~s averaging time. These results further establish future MCF networks as a promising platform for ultrastable frequency transfer.

\begin{figure*}[ht]
\centering\includegraphics[width=15.5cm]{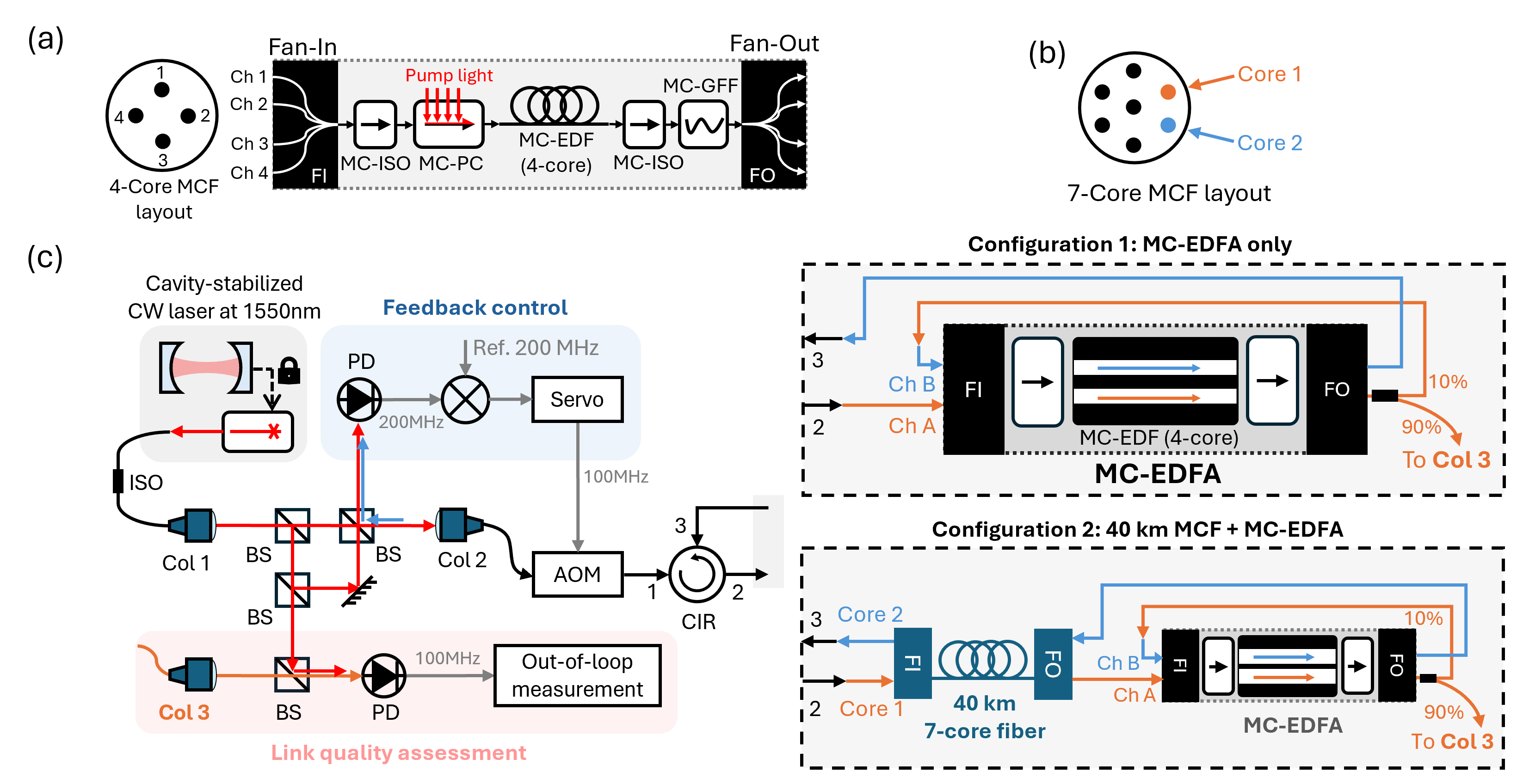}
\caption{(a) Left: core-layout and labeling scheme of the 4-core fibers in the amplifier. Right: simplified illustration of the amplifier configuration. MC-ISO: multicore isolator. MC-PC: muticore pump combiner. MC-EDF: multicore erbium-doped fiber. MC-GFF: multicore gain-flattening filter. (b) Core-layout of the 40~km 7-core fiber. (c) Schematic of the Doppler-canceled link setup. Col: collimator. BS: beam splitter. PD: photodetector. AOM: acousto-optic modulator. CIR: fiber circulator. FI/FO: fan-in fan-out devices.}
\label{Setup}
\end{figure*}

\section{Experiments}
The MCF amplifier used in our experiments is a core-pumped 4-core EDFA designed for telecom networks \cite{Wakayama:22}, with each core channel designated as Ch 1-4. A simplified configuration of the MC-EDFA is shown in Fig. \ref{Setup}(a). The amplifier consists of a 14~m long section of erbium-doped 4-core fiber, with each of the cores pumped independently at a wavelength of 980~nm through side-polished couplers. This allows individual control of each core and removes the need for pump couplers with dissimilar signal paths. The amplifier is packaged with MCF isolators at each end, such that the light in all cores propagates in the same direction. Importantly, this arrangement does not prevent assessment of the amplifier’s core-to-core noise correlation, though it does introduce excess SC fibers to route the signal through multiple cores. We note that MCF amplifiers where neighboring cores operate in opposite directions have been demonstrated \cite{Wada2025-yr}, and would simplify the link. To direct the forward and return light into separate cores for fiber link stabilization, we used a pair of fan-in-fan-out (FI/FO) devices to interface between SC fibers and the MCF section. Furthermore, using FI/FOs allows us to study the amplifier separately, and, as discussed later, to connect to a 7-core fiber to create a longer MCF link. Throughout the experiments, we operate each amplifier channel at a pump current of 400~mA (120~mW to 144~mW depending on the channels), which is typical for telecom operation. The corresponding 15~dB of gain exceeds the one-way transmission loss of the fiber and optical components.

Since each core in the MCF is pumped by a separate 980~nm source, pump power fluctuations are uncorrelated among channels, leading to uncorrelated phase shifts on the amplified outgoing and return 1550~nm signals. Thus, prior to building a fully stabilized link, we characterized the differential phase shifts due to pump power fluctuations. The magnitude of this effect depends on whether the amplifier operates in the pump-power saturation regime and potentially on the uniformity of the doping concentration across the cores. To estimate its impact on long-term phase stability of the amplified signals, we tracked the optical power fluctuations of the 980~nm pump sources in the time domain and converted them into phase fluctuations using a measured conversion factor. The conversion factor was measured by splitting stable 1550~nm light into two branches, amplifying through two different channels—one with its pump laser power externally modulated and the other kept stable—and then recombining the amplified signals on a photodetector to track the relative phase change between the signals over time (see Supplement 1 for details). 

\begin{figure}[h]
\centering\includegraphics[width=0.75\linewidth]{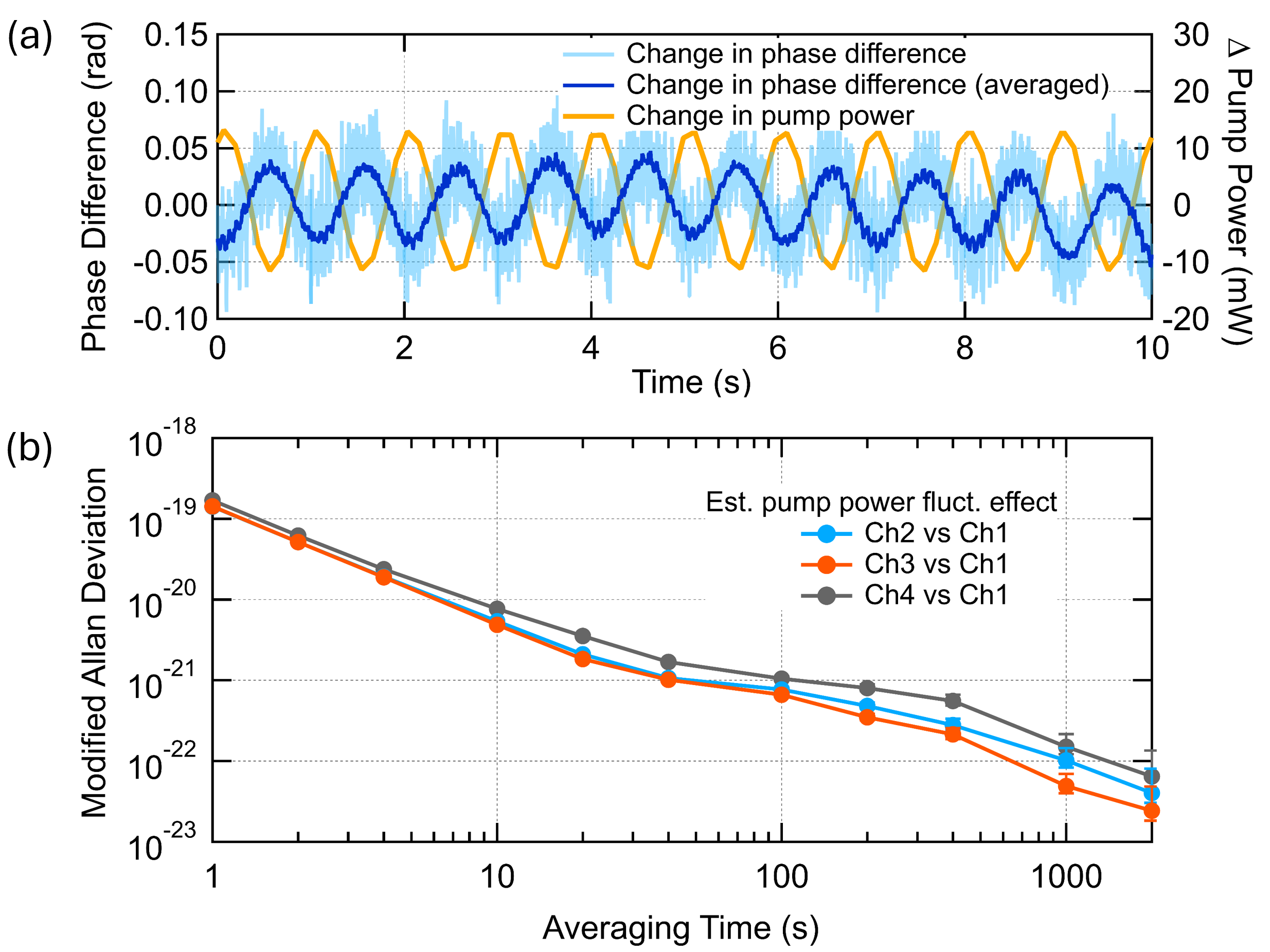}
\caption{(a) Change in pump power under external modulation and the corresponding change in relative phase between the amplified signals of the two amplifier channels. (b) Fractional frequency instability limit estimated from the pump power fluctuation effect for each amplifier channel.}
\label{AmpOnly}
\end{figure}

A representative measurement result is shown in Fig. \ref{AmpOnly}(a), displaying the pump power variation on Ch~1 under a 1~Hz modulation and the corresponding relative phase change between the amplified signals of Ch~1 and Ch~2. The pump-power-fluctuation-to-phase-change conversion factor was obtained from the ratio between the optical power change of the pump and the corresponding phase change between the amplified signals, and was measured to be around $3 \times 10^{-3}$ radian/mW for all amplifier channels at 400~mA of pump power. The estimated limitation due to these uncorrelated pump power fluctuations is shown in Fig. \ref{AmpOnly}(b) in the form of the modified Allan deviation (MDEV). The measured fractional frequency instabilities are below $10^{-20}$ for a 10~s averaging time and between $1.5\times10^{-22}$ and $5\times10^{-23}$ at 1000~s. At such small values, this effect is not expected to limit the amplifier’s performance in ultrastable frequency transfer, and thus using a common pump laser for both channels is not necessary.

More generally, there are various other factors that can affect the correlation level between cores on an MC-EDFA, such as strain along the MCFs and, more obviously, the length of non-common SC fiber paths in the FI/FOs. An evaluation of the overall correlation levels between the individual amplifier channels is discussed in more detail in Supplement 1.

To further assess the amplifier, we set up two separate noise-canceled link experiments. The first configuration stabilized the MC-EDFA alone to evaluate its capability. The second configuration stabilized a 40 km-long MCF link together with the MC-EDFA to determine whether the amplifier limits the performance of longer links. A schematic of the experimental setup for both configurations is shown in Fig. \ref{Setup}(c). In both configurations, light from a cavity-stabilized CW fiber laser at 1550 nm was sent into a free-space interferometer. Part of the light served as the local reference for error signal generation and out-of-loop performance assessment, while the rest was coupled back into fiber, frequency shifted by 100 MHz with an acousto-optic modulator (AOM), and sent through a fiber circulator into the fiber link under test. The input optical power to the link, measured after the circulator, was \textasciitilde{}0.1~mW for configuration 1 and \textasciitilde{}1.4~mW for configuration 2.

\begin{figure}[h]
\centering\includegraphics[width=0.8\linewidth]{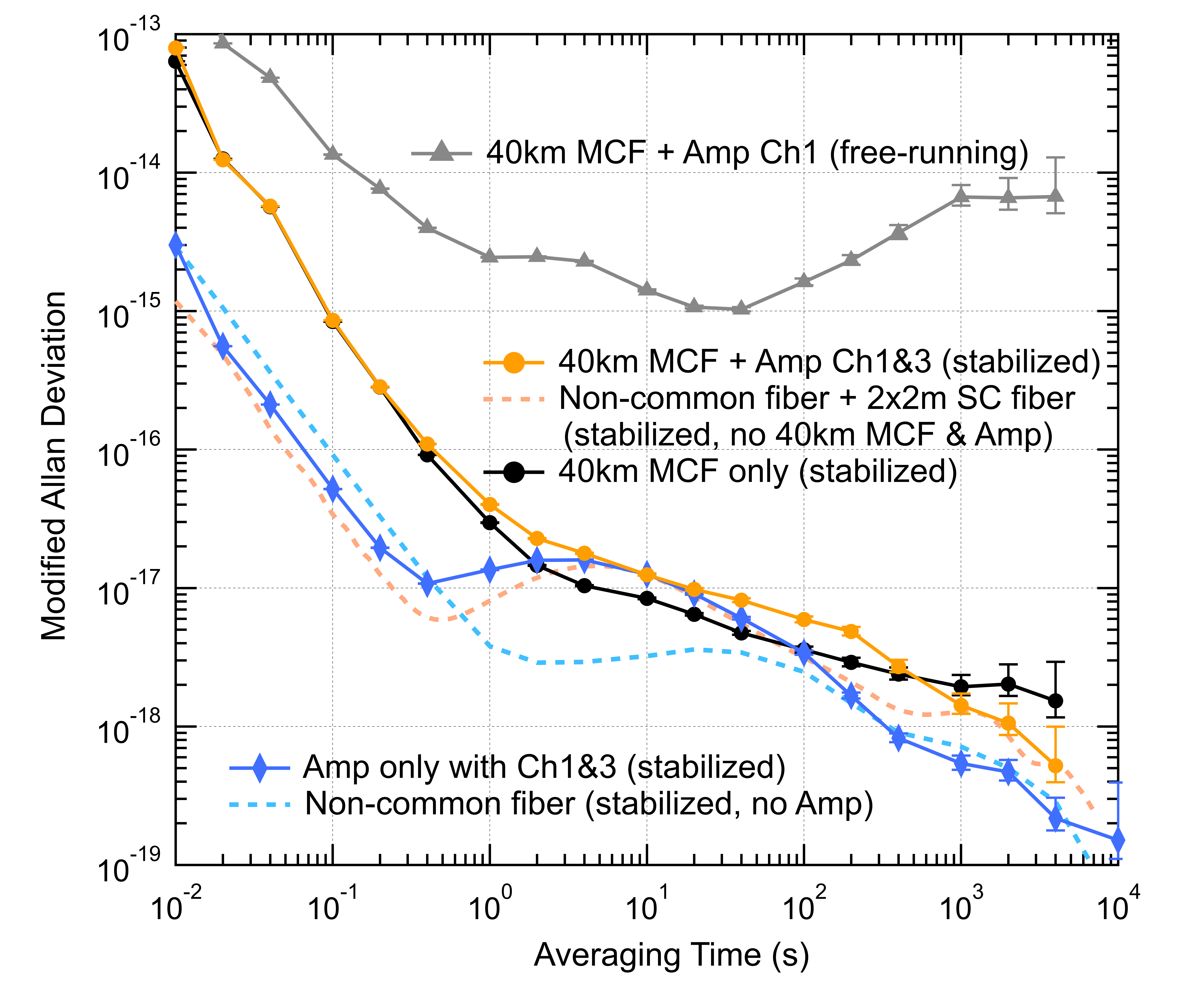}
\caption{Modified Allan deviation characterizing the long-term performance of the amplifier alone and with the 40~km 7-core fiber.}
\label{MCFandAmp}
\end{figure}

Any one of the four available amplifier channels can be used for outgoing or return light. We designate “Ch~A” as the channel used for the outgoing light, and “Ch~B” for the return light. In configuration 1, the outgoing light after the fiber circulator was launched directly into Ch~A of the amplifier. After amplification, a 90/10 fiber coupler split the signal, sending 10\% of the power into Ch~B. The remaining 90\% was used for link quality assessment by interfering it with the reference light on a photodetector, generating a 100 MHz signal for stability analysis. The signal at the output of Ch~B was routed back to the fiber circulator and passed through the AOM again before entering free space to beat with the reference light on another photodetector, producing a 200~MHz beatnote that carries the phase information of the noisy link. The 200~MHz signal was then mixed with a stable local 200~MHz signal to generate an error signal for noise compensation. The phase correction was applied by adjusting the frequency of the transmitted light using the same AOM. After the link was stabilized, the phase of the out-of-loop 100~MHz signal was tracked for 40,000~s and used to calculate the fractional frequency instability of the link. Three different combinations of channel pairs were tested. The input channel, Ch~A, was kept as Ch~1 and the return channel, Ch~B, was switched between Ch~2, Ch~3, and Ch~4. The measured fractional frequency instability for the combination of Ch~1 and Ch~3 is plotted in Fig. \ref{MCFandAmp} in the form of MDEV (blue solid trace). The MDEVs of the other combinations exhibit similar performance and are included in Supplement 1. The measured MDEV is $5.4\times 10^{-19}$ at 1000~s averaging time and continues averaging down for longer times, reaching ~$1.5\times10^{-19}$ at 10,000~s averaging time.

We note that, in order to connect the amplifier and the fiber components, approximately 1.5~m of additional SC fibers were added outside the amplifier package (shown in Fig. \ref{Setup} as the orange and blue lines), adding uncorrelated noise between the forward and return path. To assess the effect of the added fibers on the stability of the link, we measured the stability of the link while bypassing the amplifier, leaving only the fiber circulator, fiber coupler, and SC patch cords used for amplifier connection. As expected, environmental changes in the lab, such as temperature, strongly influence the measured stability, particularly at averaging times $ >100$~s. Therefore, performance can vary throughout the day and whenever the wrapping or placement of the external fibers were changed during optical connection adjustments. A worst-case scenario is shown as the blue dashed trace in Fig. \ref{MCFandAmp} for reference. The result suggests that the long-term stability of the stabilized amplifier channel measurements is likely limited by the added non-common SC fibers. The bump observed on the locked amplifier measurement (blue solid trace) between 0.4~s and 100~s is likely caused by the amplifier package itself, most probably due to the non-common SC fiber pigtails in the FI/FO devices of the amplifier, which totals approximately 4~m for a single pass.

In configuration 2, a 40~km-long, weakly-coupled, 7-core fiber \cite{Hayashi2012-np} was added before the amplifier. The fiber was spooled and has 2~m-long FI/FO devices installed at both ends. The single-pass loss of the 40~km fiber was measured to be about 11~dB. Combined with the amplifier, the total one-way gain of the link is about 4~dB. The interferometer configuration remained the same as in configuration 1, except that in this case, the light exiting port 2 of the fiber circulator was sent directly into one of the cores of the 7-core fiber (labeled as core~1), and then into Ch~A of the amplifier. The fiber routing at the amplifier was similar to that in configuration 1, except that the amplified light from the output of Ch~B was directed into the opposite end of another core of the 7-core fiber (labeled as core~2). In this way, the light traveling in core~2 propagated in the opposite direction as to that of core~1. The two selected cores of the 7-core fiber were chosen because they exhibited the highest correlation level among all core pairs \cite{nakamura2025subfem}. Additionally, the non-common SC fibers used for connection purposes in this configuration was slightly longer than those used in configuration 1, totaling about 2~m for a single pass.

We ran 6-hour duration measurements for two combinations of the amplifier channels. To establish a benchmark for analysis purposes, we also studied the performance of the stabilized 40~km MCF alone without the amplifier. Another stabilized non-common fiber measurement similar to that described in configuration 1 was also performed that bypassed both the 40~km MCF and the amplifier and added two pairs of 2~m-long SC fiber to simulate the effect of the FI/FOs. 

Results are shown in Fig. \ref{MCFandAmp}. The fractional frequency instability of the free-running amplified link (grey curve) was $7\times10^{-15}$ at a 1000~s averaging time, reducing to $1.4\times10^{-18}$ when the feedback was engaged (solid orange curve). As no significant difference in performance was found between amplifier channels, only the result for the Ch~1 and Ch~3 combination are plotted here (see Supplement 1 for additional results). Comparing the stabilized amplified link with the reference measurements without the amplifier (black) and with only the non-common fibers (dashed orange), we observed that, between 1~s and 300~s averaging times, the link stability is most likely limited by the non-common SC fibers, including those in the FI/FOs. For averaging times above 1000~s, both the non-common SC fibers and the correlation between the cores of 40-km MCF appear to set the stability limit, as their performance varies with environmental conditions. If at all, the packaged amplifier contributes at averaging times from 2~s to 100~s. In an ideal scenario where most of the links consist of MCF segments spliced together, minimizing the number of FI/FO devices and non-common connection fibers, the limitation resulted from the non-common SC fibers could be significantly reduced.

From repeated measurements, we observed that the non-common fiber measurements consistently show a downward trend beyond the 1000~s region, whereas the measurements obtained from stabilizing only the 40~km MCF exhibit larger variations over the same region. This behavior can likely be attributed to temporal changes in the correlation level between the MCF cores. It is worth noting that the 40~km fiber spool is located in a laboratory environment where the temperature and humidity are more tightly controlled, thus resulting in a lower free-running noise than that of a deployed fiber link \cite{Foreman2007}. However, the bare, unjacketed spool could be more susceptible to variations in environmental factors such as temperature and pressure change. When such change occurs, the entire 40~km optical path experiences it simultaneously, which can induce larger and more complex strain variations along the spool, resulting in greater overall changes in phase correlation between cores. Thus, we expect less change in the correlation between the cores for a deployed MCF link, as demonstrated by a deployed 25~km long 4-core fiber \cite{Hoghooghi2025-st}, in which the relative fractional frequency instability between cores reaches $1\times 10^{-19}$ at a 2000~s averaging time.

\section{Conclusion}
We have demonstrated that an MC-EDFA can support a fractional frequency instability of $5\times10^{-19}$ at a 1000~s averaging time, which could be improved by reducing the length of non-common SC fibers. When incorporated into a 40-km 7-core fiber link, the stabilized system achieved a frequency instability of $1.4\times10^{-18}$ at 1000~s, limited by the correlation level between the cores of the 7-core fiber. These results show that essential telecom components based on weakly-coupled MCFs do not pose a barrier to ultra-stable frequency transfer.

In our experiments, although the MC-EDFA was designed for unidirectional operation, a return signal for link stabilization can be conveniently acquired using FI/FOs. As we showed here, any non-common SC fiber added to the link, such as the one from FI/FOs, can degrade link performance. Hence, we expect an improved link performance with a true bidirectional MC-EDFA without FI/FOs. Moreover, operating an MCF link where neighboring cores operate in counterpropagating direction has proven to be extremely beneficial, as it can greatly suppress inter-core crosstalk and increase transmission distance \cite{Soma2024-or, Takeshita2024-jx}. As a result, FI/FO-less bidirectional MC-EDFAs are being actively investigated, and recent progress appears promising \cite{Wada2025-yr}.

In a deployed MCF link, isolators can be integrated for individual cores around each MC-EDFA where different cores operate in different directions. With separate gain control on the separate cores, the optical power of the outgoing and return light can be independently tailored as needed. Thus, combining independent gain control with isolation that mitigates back-reflections and spurious lasing, ultrastable links can be developed with net gain. In our experiment, we achieved an overall one-way gain of 4~dB over a 40 km link, and this gain could be increased further if needed. This is contrast to current bidirectional links over SC fiber, where the net link loss can be high \cite{Predehl2012-df}. With the frequency stability demonstrated here, future MCF networks could serve as a promising platform for time and frequency distribution.

\vspace{2em} 

\noindent\textbf{Funding.} National Institute of Standards and Technology.

\medskip

\noindent\textbf{Acknowledgment.} We thank Esther Baumann and Tara Fortier for helpful comments on the manuscript.

\medskip

\noindent\textbf{Disclosures.} The authors declare no conflicts of interest.

\medskip

\noindent\textbf{Data availability.} Data underlying the results presented in this paper are not publicly available at this time but may be obtained from the authors upon reasonable request.

\medskip

\noindent\textbf{Supplemental document.} See Supplement 1 for supporting content.

\newpage
\bibliographystyle{unsrt}
\bibliography{References}

\newpage
\clearpage
\appendix 

\setcounter{section}{0}
\setcounter{figure}{0}
\setcounter{table}{0}
\setcounter{equation}{0}

\renewcommand{\thefigure}{S\arabic{figure}}

\begin{center}
    \vspace*{2em}
    {\Large \textbf{Supplemental Material:}} \\
    \vspace{0.5em}
    {\Large \textbf{Fiber Link Stabilization with a Multicore Fiber Amplifier}} \\
    \vspace{2em}
\end{center}

\section{Cavity-stabilized laser}
The laser used in this experiment is a 1550~nm fiber laser stabilized to an ultra-stable Fabry-Perot cavity in vacuum. The Fabry-Perot cavity is of cylindrical shape with a 25~mm-long spacer made of ultra-low-expansion glass and mirror substrates made of fused silica glass. Its Allan deviation is about $5\times10^{-15}$ at 1~s with a \textasciitilde{}1~Hz/s linear drift.

\section{Amplifier inter-channel phase correlation measurements}
\begin{figure}[htbp]
\centering
\includegraphics[width=0.8\linewidth]{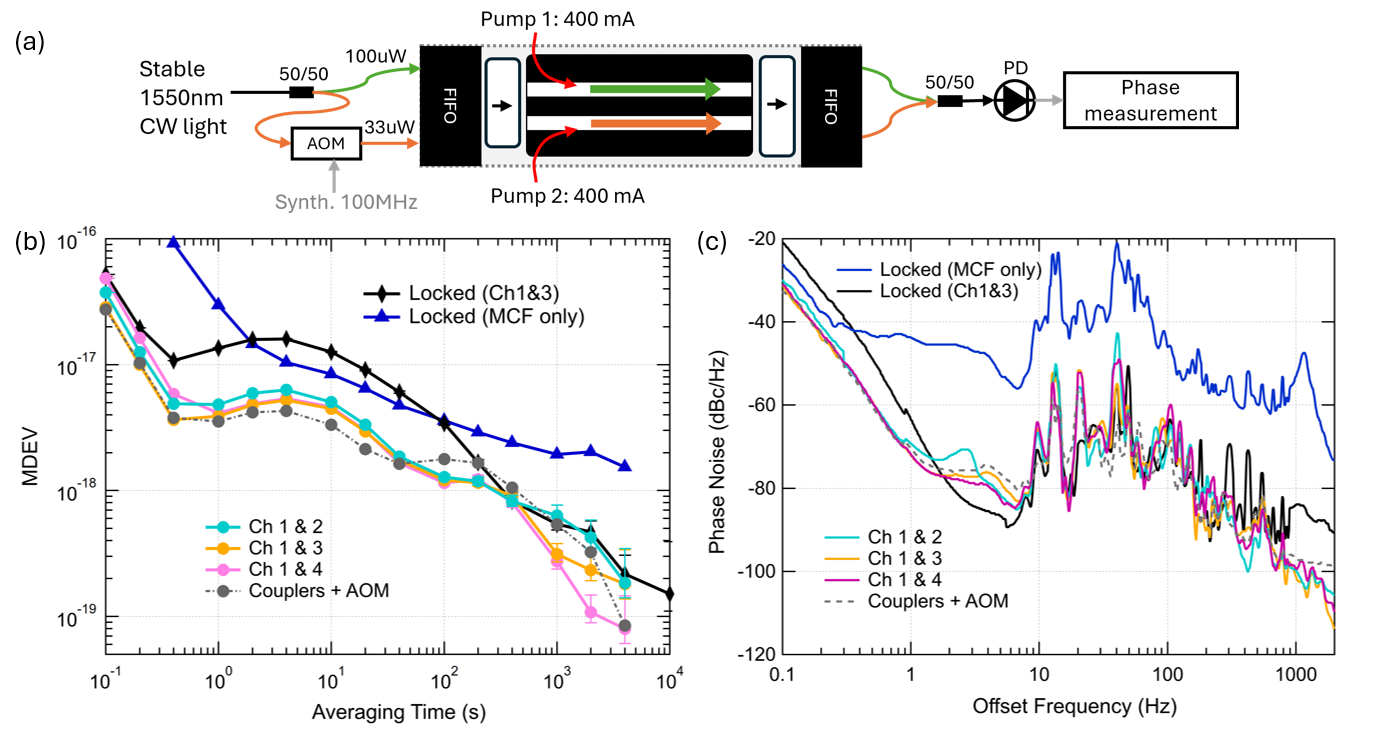}
\caption{(a) Schematic illustration of the inter-channel correlation measurements of the MCF amplifier. (b) Modified Allan deviation estimated from the inter-channel phase correlation measurements. (c) Phase noise spectrum of the measured correlation level.}
\label{fig:AmpCorr}
\end{figure}

As discussed in the main text, we assessed the phase correlation between different channels of the amplifier to set a limit for the achievable stability of a Doppler-canceled link. The schematic illustration of the experiment is shown in Fig. \ref{fig:AmpCorr}(a). The stable continuous-wave (CW) light at 1550~nm was split into two branches. One branch was sent directly into one of the amplifier channels. The other branch was frequency shifted by 100~MHz through a fiberized AOM and sent into another amplifier channel. The pump currents for both channels were maintained at 400~mA. At the output, the amplified light from the two channels were combined with a fiber coupler and sent to a photodetector. The phase change over time on the 100~MHz beatnote is recorded and analyzed. 

Assuming the phase change experienced along one channel is represented as $\Phi_A = \phi$ and the the phase change experienced by the other channel is represented as $\Phi_B = \phi + \delta$, where $\phi$ represents the common phase change experience by both channels and $\delta$ represents the non-common phase change between the channels, the phase change we measured from this experiment on the 100~MHz beatnote should be $\delta$. However, if we use the two channels for a Doppler-canceled link, where the light first passes one channel and comes back from the other channel to compare with the local reference for correcting the noise added to the link, the phase change at the output of the first channel after the correction should be $\Phi_A - \frac{1}{2}(\Phi_A +\Phi_B) = \phi - \frac{1}{2}(2\phi +\delta) = -\frac{\delta}{2}$. Therefore, from this simplistic estimation, the achievable phase fluctuation of such a Doppler-canceled link should be half of the phase change measured on the 100~MHz beatnote, corresponding to a 6~dB difference for power spectral density.

The time-domain result for the estimated Doppler-cancellation-limit is shown in Fig. \ref{fig:AmpCorr}(b) in the form of the modified Allan deviation (MDEV), and the frequency-domain result is shown in Fig. \ref{fig:AmpCorr}(c) in the form of single-side-band phase noise. 

The blue traces are the measured MDEV and phase noise for the stabilized 40~km MCF link, displayed here for reference purpose only. They clearly show that we should be able to incorporate the amplifier into the 40~km without affecting the link performance, since the projected MDEV and phase noise are well below what is measured for the stabilized 40~km link alone. We also ran the same correlation measurement without the amplifier, measuring the phase change on the 100~MHz beatnote with only the couplers and the AOM. The results are the grey dashed traces that largely overlap with the inter-channel correlation measurements, indicating that the non-common fiber components in the experimental setup clearly set a measurement floor, and a more accurate estimate of the inter-channel correlations are not revealed. From the MDEV results, we can see that the relative phase change between the channels is small, leading to fractional frequency instability of about $1.3\times 10^{-18}$ at 100~s averaging time and below $10^{-18}$ at 1000~s. These results indicate that the amplifier should be able to support a Doppler-canceled link with at least a $10^{-18}$ level performance. 

The black traces are the measured MDEV and phase noise of the stabilized amplifier alone using Ch~1 and Ch~3, displayed here for reference purpose as well. The phase noise of the stabilized amplifier aligns with our projected phase noise of the inter-channel correlation measurements for offset frequencies higher than 10~Hz, most likely limited by seismic and vibration effects. For longer times, as we discussed in the main text, the performance is limited by the non-common SC fibers.

\section{Instability from pump power fluctuation}
\begin{figure}[htbp]
\centering
\includegraphics[width=0.8\linewidth]{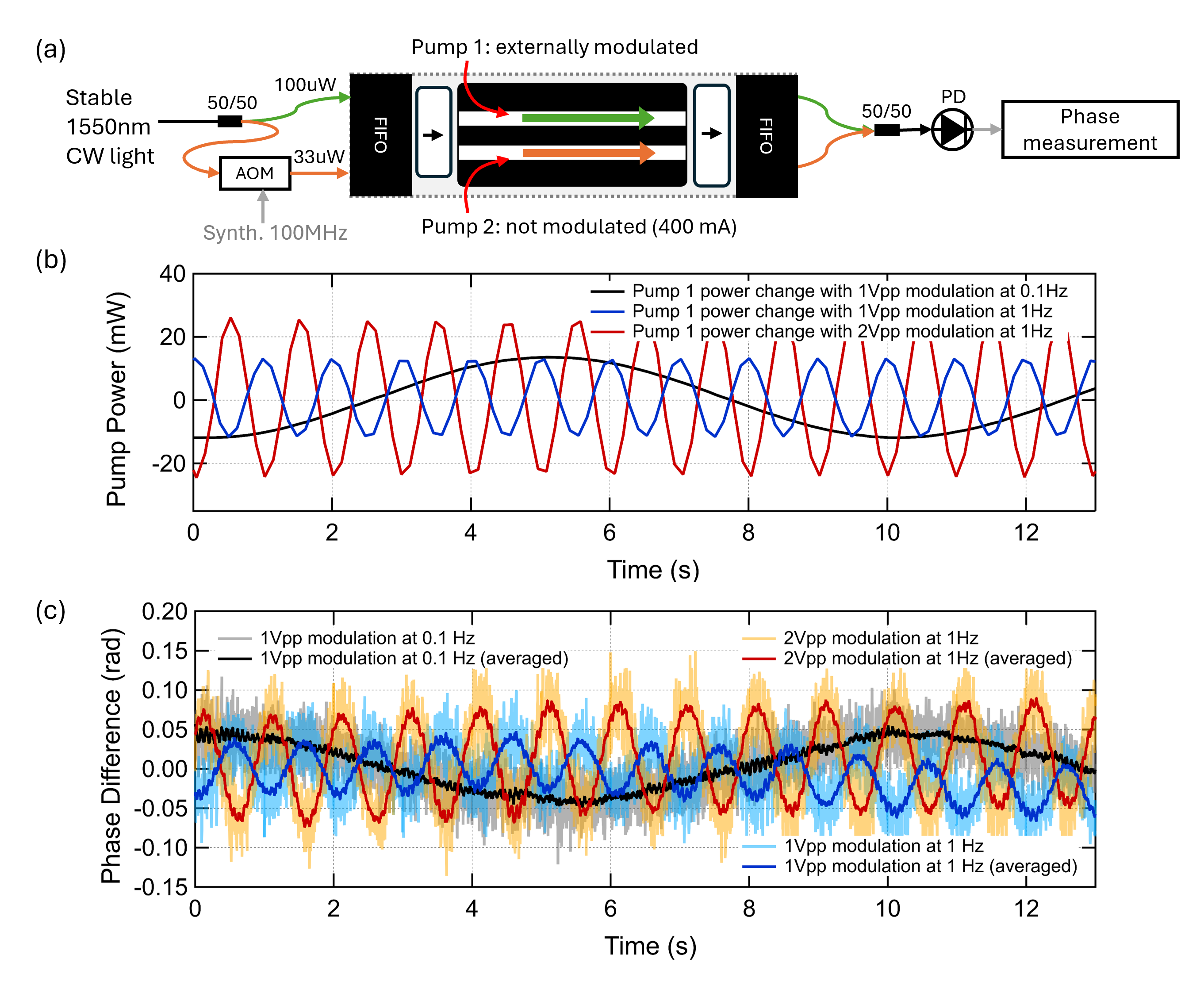}
\caption{(a) Schematic illustration of the pump power vs. phase change measurements of the MCF amplifier. (b) The measured pump power change and (c) the corresponding phase change.}
\label{fig:PumpPowerVsPhase}
\end{figure}

As stated in the main text, the power fluctuation on the pump lasers can limit the correlation level between the amplifier channels. Since each channel of the amplifier is pumped by a separate 980~nm laser, the power fluctuations on the lasers are uncorrelated between channels, thus leading to non-common phase variations on the amplified signals.

To quantify this phenomenon, we measured the conversion factor between the optical power change of the pump laser and the corresponding phase change on the amplified light. The schematic experimental setup is shown in Fig. \ref{fig:PumpPowerVsPhase}(a). The experiment is similar to that described in section 2 of this document, except that this time one of the pump lasers is externally modulated by a sine wave, and the other pump laser remains stable at a pump current of 400~mA. With the modulation on, the corresponding phase difference between the two amplifier channels is recorded. Measurement results for Ch~1 and Ch~2 of the amplifier are shown in Fig. \ref{fig:PumpPowerVsPhase}(b) and (c). Figure \ref{fig:PumpPowerVsPhase}(b) shows the change in power of the Ch~1's pump laser under modulation with two different modulation depths, 1~Vpp and 2~Vpp, and two different modulation frequencies, 1~Hz and 0.1~Hz. Fig. \ref{fig:PumpPowerVsPhase}(c) shows the measured phase variation under the corresponding modulations on a 1550~nm carrier. As the phase difference measurement can experience nonlinear drifts overtime caused by the non-common fiber path, we manually select intervals with linear drift trend and perform linear drift removal to recover the sinusoidal response. Comparing Fig. \ref{fig:PumpPowerVsPhase}(b) and Fig. \ref{fig:PumpPowerVsPhase}(c), we see that the phase change is proportional to the pump power change under 1~Hz modulation, with about 23~mW of pump power change corresponding to about 0.07~radian phase change on 1550~nm carrier, yielding a conversion factor of $3 \times 10^{-3}$ radian/mW. Under the slower 0.1~Hz modulation, the resulted phase change is about 0.9~radian for 25~mW of change in pump power, leading to a conversion factor of $3.6 \times 10^{-3}$ radian/mW. No significant difference is observed when the modulation is switched to the other channel (Ch 1’s pump laser remains at 400 mA, while Ch 2’s pump laser is modulated). We tested three pairs of channels total, with the other two pairs being Ch~1 and 3 and Ch~1 and 4. The measured conversion factors are in the range of $2$ to $4 \times 10^{-3}$ radian/mW, and we pick $3 \times 10^{-3}$ radian/mW for calculation in the main text.

We then tracked the optical power fluctuations of the 980~nm pump sources in the time domain and converted them into phase fluctuations using the measured conversion factor. As discussed in Section 2, we used half of the converted phase fluctuations to calculate MDEV, which set a limit for the stability of the Doppler-canceled link.

(We note that the corresponding optical power for each channel at 400~mA pump current is different. For all measurements, the pump power is 130~mW, 120~mW, 136~mW, and 144~mW for Ch~1, Ch~2, Ch~3, and Ch~4 respectively.)

\section{Fractional frequency instability with other amplifier channel combinations}
\begin{figure}[htbp]
\centering
\includegraphics[width=0.7\linewidth]{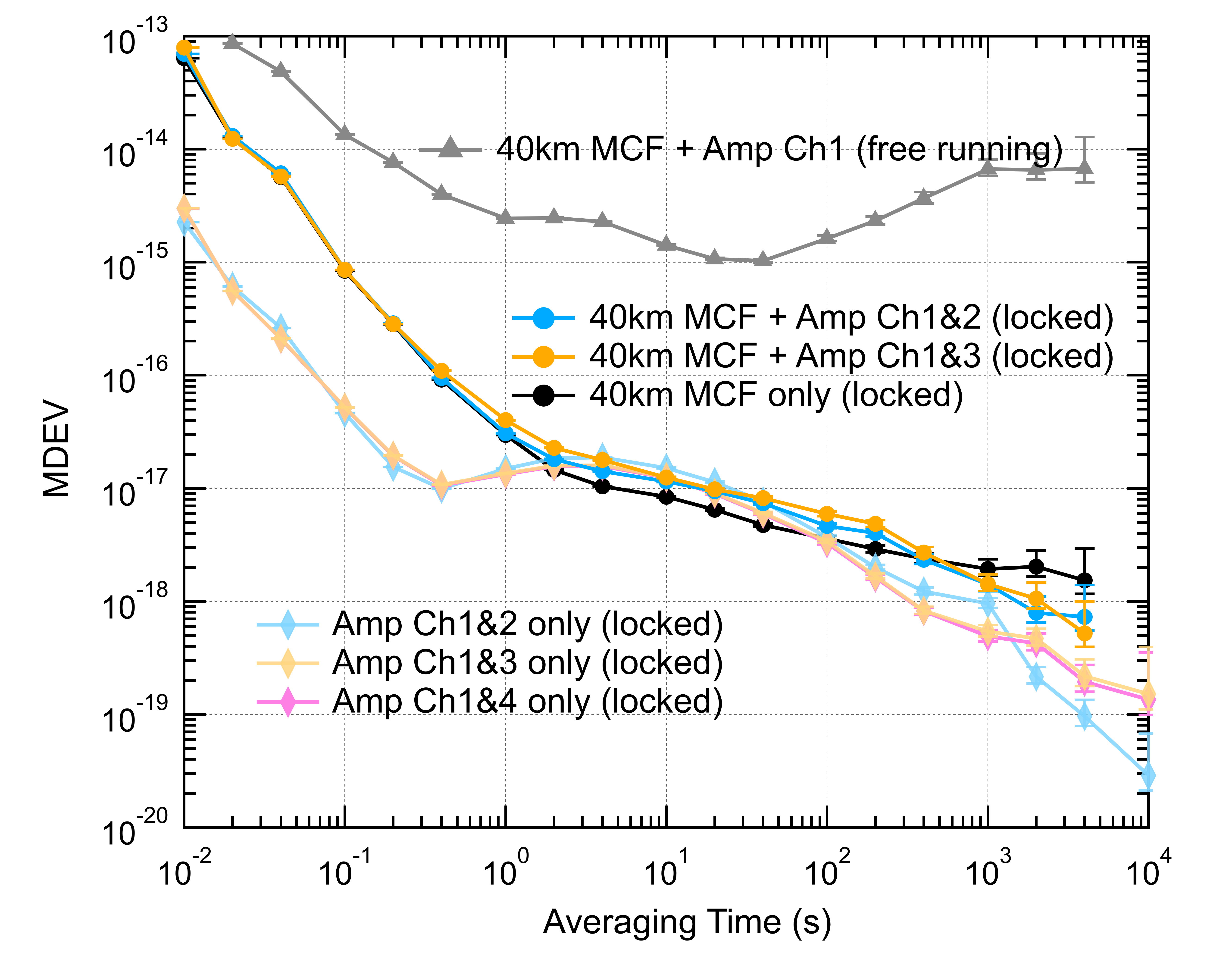}
\caption{Modified Allan deviation of the stabilized links.}
\label{fig:MDEVSummary}
\end{figure}
The measured fractional frequency instabilities with different amplifier channel combinations are plotted in Fig.~\ref{fig:MDEVSummary}. For the configuration 1 measurements (stabilizing the amplifier alone), three combinations of amplifier channels were tested, which were Ch~1 with Ch~2 (light blue trace with rhombus), Ch~1 with Ch~3 (light orange trace with rhombus), and Ch~1 with Ch~4 (pink trace with rhombus). Overall, the measured MDEVs are below $1\times 10^{-18}$ at a 1000~s averaging time and continue averaging down for longer times, reaching ~$1.5\times10^{-19}$ or lower at a 10,000~s averaging time. As discussed in the main text, the MDEV after 1000~s averaging times are limited by the non-common SC fibers in the setup. For the configuration 2 measurements (stabilizing the 40~km 7-core fiber with the amplifier), two combinations of the amplifier channels were tested, which were Ch~1 with Ch~2 (blue trace with dots) and Ch~1 with Ch~3(blue trace with dots), since they exhibited slightly different behavior in the configuration 1 measurements. No significant difference in performance was observed between the two combinations.

\section{Phase noise spectra of the stabilized links}
\begin{figure}[htbp]
\centering
\includegraphics[width=0.7\linewidth]{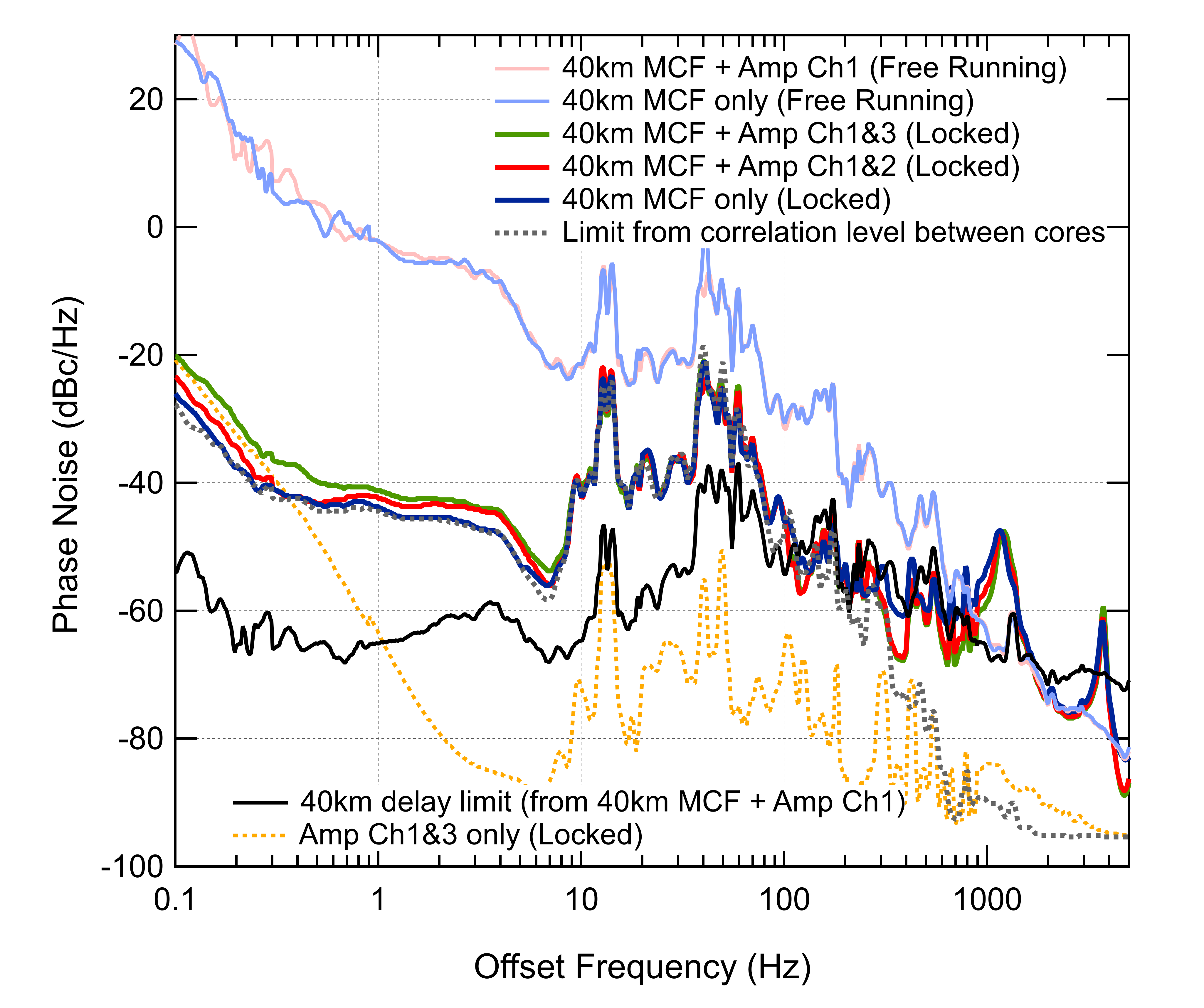}
\caption{Phase noise spectrum of the stabilized links.}
\label{fig:PNSummary}
\end{figure}
As the long term stability of the stabilized links are discussed in the main text and section 4 of this document, here we display phase noise power spectra, corresponding to the short-term behavior in Fig. \ref{fig:PNSummary}. The top two traces (pink and light blue) correspond to the free-running noise of the 40~km MCF link and the free-running noise of the 40~km MCF with the MCF amplifier respectively, showing that the free-running noise is dominated by the 40~km MCF and the amplifier adds insignificant noise to the link, which fits our expectations. The green and red traces correspond to the noise measured on the stabilized 40~km MCF link with the MCF amplifier, and the darker blue trace corresponds to the noise on the stabilized 40~km MCF link without the amplifier. The grey dashed trace corresponds to a correlation measurement (similar to that described in section 2)  performed between the two cores of the 40~km MCF link, which largely overlaps with the blue trace, indicating the stabilized link is limited by the inter-core correlation of the 40~km MCF, or likely the non-common SC fibers, such as the FI/FOs of the 40~km MCF. The black trace is calculated from the free-running noise and is the projected limit based on the length of the link \cite{Williams2008-gr}, which is well below the inter-core correlation limit for offset frequencies less than 100 Hz. However, this does not mean that the inter-core correlation is fundamentally incapable of achieving a similar level. As we mentioned, the correlation limit at low offset frequencies likely comes from the non-common SC fibers and components. We expect that reducing the amounts of the SC fibers can lower this limit. The yellow dashed trace is a noise measured for only stabilizing the amplifier, which shows far lower noise offset frequencies greater than 1 Hz, but with rapidly increasing noise for lower offset frequencies. This again can be explained by the existence of the non-common SC fibers.

\end{document}